\documentclass{ws-ijmpd}

\begin{document}

\markboth{Authors' Names}
{Instructions for Typing Manuscripts (Paper's Title)}

%
\catchline{}{}{}{}{}
%

\title{A ROBUST TEST OF GENERAL RELATIVITY IN SPACE  }

\author{JAMES GRABER\footnote{jgraber@mailaps.org}}

\address{Technology Assessment, Library of Congress, 101 Independence Ave. SE\\
Washington, DC 20540,
USA\\
jgra@loc.gov}

\maketitle

\begin{history}
\received{Day Month Year}
\revised{Day Month Year}
\comby{Managing Editor}
\end{history}

\begin{abstract}
LISA may make it possible to test the black-hole uniqueness theorems 
of general relativity, also called the no-hair theorems, by Ryan's method of detecting the quadrupole moment of a black hole using high-mass-ratio inspirals.  
This test can be performed more robustly by observing inspirals in earlier stages, where the simplifications 
used in making inspiral predictions by the perturbative and post-Newtonian methods are more nearly correct.  Current concepts for future missions such as DECIGO and BBO would allow even more stringent tests by this same method.  Recently discovered evidence supports the existence of intermediate-mass black holes (IMBHs).  Inspirals of binary systems with one IMBH and one stellar-mass black hole would fall into the frequency band of proposed maximum sensitivity for  DECIGO and BBO.  This would enable us to perform the Ryan test more precisely and more robustly.  We explain why tests based on observations earlier in the inspiral are more robust and provide preliminary estimates of possible optimal future observations.
 
\end{abstract}

\keywords{LISA; BBO; DECIGO; no-hair.}

\section{Introduction}	

The theme of this NASA workshop {\it From Quantum to Cosmos: Fundamental Physics Research in Space} is testing fundametal physics in space, celebrating past such accomplishments and anticipating possible 
future achievements.  One fundamental test of general relativity that apparently depends on space-based 
gravitational wave detectors for practical implementation is the test of the black-hole uniqueness theorems  (no-hair theorems) first proposed by Ryan [1].  Testing general relativity is one of the official goals of the LISA project [2,3] and includes specifically measuring the extreme-mass-ratio inspirals (EMRIs) [4] that are necessary to perform Ryan's test.  
Ryan [5] concluded that LISA could perform his test to an accuracy of order one percent with data from a favorable EMRI.
In Graber [6], we reached a similar conclusion.
  The prospects that we will be able to perform robust and accurate versions of Ryan's test have
 brightened considerably due to the recent discovery of probable intermediate-mass black holes, which  increases not only the number of  
gravitational-wave-dominated binary inspirals that are likely to be seen, but also the likelihood that we can observe them 
in the stages of the inspiral where the predictions are most robust and where the data is most likely to support precise and reliable tests.

If substantial numbers of IMBHs exist, as recently proposed [7,8,9], it will be possible to perform a greatly enhanced Ryan test with future possible space missions such as Big Bang Observer (BBO) [10,11] or DECIGO [12]. 
This is because the inspiral of a stellar-mass black hole into an IMBH falls into the most sensitive band of  BBO or DECIGO, 
where there are no interfering white dwarf binaries,
and where it will spiral through millions of cyles in less than ten years.
BBO, which is optimized to find faint gravitational waves from the big bang itself,
will be more than a thousand times as sensitive as LISA and will be able to see light IMRIs throughout the entire universe.

In this paper we briefly review the recent developments affecting our expectations of observing 
extreme- and intermediate-mass-ratio inspirals, (EMRIs and IMRIs), and
consider the eventual possibilities for performing more robust and more accurate tests of general relativity.
We point out that data from early in the inspiral has some advantages over data from later stages 
for performing robust and accurate tests.  We give order-of-magnitude estimates for the possible improvements in accuracy 
and for possible increases in the number of systems observed to indicate the potential rich harvest that awaits these 
future, more sensitive, missions to test fundamental physics in space by observing black holes 
with gravitational waves.

\section{Definition of EMRIs, Light and Heavy IMRIs.}

For simplicity supermassive black holes are defined as those greater than $10^6$ solar masses,
stellar-mass black holes as those less than 100 solar masses, and
intermediate-mass black holes (IMBHs) as those from $10^2$ to $10^6$ solar masses.

A classic EMRI is the inspiral of a stellar mass black hole into a supermassive black hole.
A heavy IMRI is the inspiral of an IMBH into a supermassive black hole.
A light IMRI is the inspiral of a stellar mass black hole into an IMBH.

\section{Short Summary of DECIGO and BBO Proposals:}

BBO and DECIGO are concepts for far more sensitive, space-based gravitational wave observatories to follow LISA.
One of the key ideas of the DECIGO and BBO proposals is to put LIGO and VIRGO technology in space.
Another key factor in these proposals is arm lengths ten times shorter than LISA, resulting in peak sensitivities at higher frequencies.
BBO (Big Bang Observer), in particular, is optimized to detect very weak gravitational waves from the Big Bang itself.
The fact that this also makes it so useful for performing Ryan's test with light EMRIs is a bonus.

The inclusion of shorter arm lengths will make BBO/DECIGO-type systems not only more sensitive than LISA, but also sensitive to different sources.  It turns out that the inspirals of light IMRIs fall right into this sensitivity band.
LISA's peak sensitivity is approximately $10^{-20}$ strain per root Herz from .003 Hz to .01 Hz.
Proposed DECIGO and BBO systems are planned to have peak sensitivity of $10^{-23}$ strain per root Hz from .1 to 1.0 Hz, i.e., about 1000 times more sensitive in a frequency band 10 to 100 times higher.
This band is ideally suited for observing the inspirals of light IMRIs.

\section{Short Summary of Testing General Relativity by Ryan's Method}

The basic observable gravitational wave form is quasi-sinusoidal with a slowly rising frequency, called a chirp.  The phase of this sinusoid ($\phi$) corresponds to twice the phase of the orbiting binary.  It can be recovered exactly by removing Doppler shifts for the appropriate direction and  referring the LISA signal to the solar system barycenter.  By matched filtering, we can determine the frequency of the chirp as a function of time with an error of less than a single cycle in the length of the filter, which can potentially be many thousands of cycles long.  This frequency evolution function (FEF) (technically $d\phi/2 \pi dt$ as a function of time) will be observed with this accuracy over tens or hundreds of thousands --or even a million or more --cycles in a typical chirp observed by LISA.  Hence the FEF will be known with an accuracy better than one part in $10^5$ or $10^6$.  This is what enables us to perform precision tests of general relativity, by comparing the observed FEF to a predicted FEF.

According to the black-hole uniqueness theorems [13-17], in general relativity the only astrophysically possible neutral black hole is a Kerr black hole, which is uniquely determined by its mass M and spin S.  General relativity predicts that the magnitude of the suitably defined quadrupole moment Q of a Kerr black hole is  $Q=S^2/M$.   If Q is not equal to  $Q=S^2/M$, general relativity is falsified.   

Ryan [1] showed  that one can determine the mass M, the spin S and the quadrupole moment Q from just the first four terms in the Taylor expansion of the FEF in the extreme-mass-ratio circular-orbit case.  Put another way, Ryan showed that if you can measure the first three terms of this series, you can predict the fourth.  Use of this decomposition of the FEF to check whether or not $Q=S^2/M$, is the test of the black-hole uniqueness theorems by Ryan's method, or the Ryan test. This is one of the easiest and cleanest tests for the correctness of general relativity, and one of the most restrictive on possible alternate theories of gravity.  In principle, one needs only three numbers (M, S, Q) for this test.  

Since the FEF is a convergent series [particularly far away from the innermost circular orbit (ISCO)],
the first four terms are generally decreasing,
and the accuracy of the test is determined by the size of the fourth (smallest) term.

Since the number of cycles is the most directly measurable feature, and the error is of the order of one cycle, 
the dominant error is of the order one over the number of cycles contributed by the fourth term.

The {\it accuracy} of the test is determined by how precisely we can {\it measure} the number of cycles 
contributed by the first four terms of this series.

The {\it robustness} of the test is determined by how precisely we can {\it predict} the number of cycles 
contributed by the first four terms of this series.

\section{Why Earlier is Better}

The lack of theoretical robustness in general relativity inspiral predictions primarily comes from  uncomputed higher-order terms [18]
and from the progressive failure of the adiabatic hypothesis and other simplifications made to compute these inspiral predictions [19].
It is well known that these errors and deviations get larger near the ISCO [20-30].
On the contrary, the unknown terms become less important and the approximations become more acurate as 
one moves earlier in the inspiral and farther from the ISCO.

Also important is how well we can isolate, observationally and theoretically, the number of cycles contributed by higher order terms, as well as the terms of the first four orders.

Due to a prefactor of order minus five,  as you move ealier in the inspiral (and away from the ISCO), the number of cycles contributed by terms of order four and less {\it increase}, whereas the number of cycles contributed by terms of order six or higher {\it decrease}.
Since the contributions of these higher order terms decrease as we move away from the ISCO,
it is easier to get an accurate measurement of the lower order terms farther from the ISCO, 
as long as there is enough frequency sweep to cleanly separate the terms of different orders.

Thus, it is more robust to measure the inspiral 
at an earlier stage, somewhat removed from the ISCO,
for two reasons:

First, the general relativity predictions are cleaner and more robust theoretically.

Second, the measurement of the contributions of the lower-order terms needed for Ryan's test are more precise.  

We will see that light IMRIs and proposed second- and third-generation missions, i.e., DECIGO and BBO,
help achieve these objectives of getting more inspiral cycles farther from the ISCO.

\section{Summary of Evidence for IMBHs}

A small number of nearby globular clusters and dwarf galaxies have shown dynamical evidence 
consistent with IMBHs [31].  A very large number of ultra-luminous X-ray sources (ULXs) have been observed, on the order of several per L* galaxy [32]. 
If  a significant fraction of ULXs are IMBHs, as now seems likely, IMBHs are approximately as numerous as L* galaxies.
For supermassive black holes, it is commonly accepted that there is one in almost every L* galaxy, 
but they are only actively emitting  X-rays about one per cent of the time.
If a similar ratio of IMBHs are active as ULXs at any time, that would imply that IMBHs are of order 100 times as numerous as L* galaxies.

Another, somewhat more speculative, line of reasoning merely assumes that IMBHs 
are approximately as numerous as globular clusters [33], 
since some dynamical evidence supports IMBHs in globular clusters, (e.g. M15 [34],and G1 [35]),
and some ULXs are associated with globular clusters. 
This also results in a ratio of IMBHs to L* galaxies of order 100 to 1.

This same type of argument can be given for dwarf galaxies in place of globular clusters.
The evidence is less firm, but the expected relative numbers are again the same within an order of magnitude.

Hereafter we assume for our optimistic estimate that IMBHs are 100 times as numerous as L* galaxies, 
with of course, large uncertainties.

Another argument for the existence of IMBHs is that almost all supermassive black hole formation scenarios pass through an IMBH stage [36].

 The simulations of IMBH formation in globular clusters suggest that it is a natural result of runaway core collapse 
and stellar collisions in the central cusp of the globular cluster. 
Many stellar-mass black holes are expected to be present and to be absorbed by the growing IMBH [37].
The formation of light IMRIs in globular clusters is highly likely in this scenario.

\section{Why Light IMRIs Give a More Precise and Robust Ryan Test}

The overall sensitivity of the Ryan test is proportional to the number of cycles of the inspiral that are observed.
The higher frequency of the BBO band and the light IMRIs, 
as compared to the LISA band and the heavy IMRIs and classical EMRIs, 
results in 100 times more cycles in the same amount of time.

As discussed in Section 3, the robustness and the accuracy of the Ryan test are greater earlier in the inspiral.
  
The classical EMRIs and the heavy IMRIs begin to get lost in the white dwarf binary confusion noise as one 
moves away from the ISCO.  The light IMRIs have at least two extra decades of frequency sweep before they hit that limit.
Also, as they accumulate cycles more than 100 times faster, their measurement is also less likely to be impacted by the mission duration limit.
Hence the light IMRIs with BBO or DECIGO 
are likely to permit a very substantially more robust and precise measurement than the EMRIs (or heavy IMRIs) and LISA.

\section{Conclusion}
We have briefly explained why earlier inspiral data is more theoretically robust. 
It contains a greater number of total cycles and a higher number of cycles per octave of frequency sweep.
It also contains a higher ratio of predicted cycles to unpredicted cycles.
Light IMRIs in the .1 Hz band are likely to give a very robust and precise Ryan test when BBO or DECIGO flies.

\end{document}